# On the carrier transport and radiative recombination mechanisms in tunneling injection quantum dot lasers

V. Mikhelashvili[1], S. Bauer[2], I. Khanonkin[1], O. Eyal[1], G. Seri[1], L. Gal[1], J.P. Reithmaier[2], G. Eisenstein[1]

[1] Electrical Engineering department and Russel Berrie Nanotechnology Institute, Technion, Haifa 32000, Israel
[2] Institute of Nanostructure Technologies and Analytics, Technische Physik, CINSaT, University of Kassel, 34132 Kassel, Germany

**Abstract**

We report temperature-dependent current–voltage ($I - V - T$) and output light power-voltage or current ($P - V - T$) or ($P - I - T$) characteristics of 1550 nm tunneling injection quantum dot (TI QD) laser diodes. Experimental data is accompanied by physical models that distinguish between different current flow and light emission mechanisms for different applied voltages and temperature ranges. Three exponential regimes in the $I - V$ characteristics were identified for low bias levels where no optical radiation takes place. At the lowest bias levels, the diffusion-recombination mechanism based on the classical Shockley-Reid-Hall theory dominates. This is followed, at low and near room temperature, by a combination of weak tunneling and generation-recombination, respectively. In the third exponential region, for all temperatures carrier transport is dictated by strong tunneling, which is characterized by a temperature-independent slope of the $I - V$ curves and a variable ideality factor. The $I - V$ results were compared to a conventional QD laser in which the current flow mechanisms of the first and third types are absent, which clearly demonstrates the key role played by the TI layer.

In the post-exponential voltage range, when the diodes are in the high injection regime, the characteristics of the two types of diodes are identical. The typical behavior at the threshold current, where the output power increases fast has a clear signature in the $I - V$ characteristics. Finally, an analytical quantitative relationship is established between the light output power and the applied voltage and current as well as the carrier density participating in radiative recombination.

## I. Introduction

The concept of tunneling injection (TI) in semiconductor lasers [1] has been proposed and demonstrated close to two decades ago in order to increase efficiency, lower threshold currents and widen the modulation bandwidth. Lasers using TI were demonstrated first in quantum well (QW) [2] and then in quantum dot (QD) lasers [3-5]. Every TI laser includes a carrier injection reservoir (typically a QW placed close to the emitting layer) from which cold carriers are injected directly to the lasing energy states, thus avoiding the typically slow cascaded relaxation of hot carriers from high energy states. A proper TI process ensures therefore, in addition to low threshold currents and high efficiency, a reduced junction temperature and high-speed modulation capabilities. Two possible TI processes take place simultaneously in QD lasers [6, 7] non-resonant TI which is mediated by phonons and resonant TI where the lowest level of the tunneling injector is energetically aligned with an excited state of the QD forming a hybrid state from which carriers relax to the QD ground state.

The details of the TI process were modeled using various approaches [8-10]. The two TI processes were identified directly using multi wavelength pump probe characterization in a QD optical amplifier [7]. The non-resonant TI process and the role of phonons were observed experimentally using pump-probe measurements [5], while hybrid state dynamics were mapped out using photoreflectance and photoluminescence [11], from which the carrier transfer efficiency from reservoir to QD was quantified.

In this paper we address the details of carrier transport in a TI QD laser using static, temperature dependent electrical and electro optical characterizations. Similar measurements reported previously for QD lasers [12], identified specific carrier transport mechanisms at different applied bias regions. For the TI QD laser, we identify five transport regimes seen in the current-voltage characteristics ($I - V$) and their corresponding power exponent parameter dependence on voltage ($\alpha - V$, where $\alpha = \frac{d[\ln(I)]}{d[\ln(V)]}$) [12]. These regimes are also clearly seen in the optical output power-voltage ($P - V$) and $\gamma - V$ characteristics, where $\gamma = \frac{d[\ln(P)]}{d[\ln(V)]}$. Similarly, they are observable in the more conventional $P - I$ and $\beta - I$ characteristics, with $\beta = \frac{d[\ln(P)]}{d[\ln(I)]}$. $P - V$ characteristics are rarely used in diode laser research but are advantageous over the common $P - I$ characteristics when intricate details of carrier transport mechanisms are investigated. The $I - V$ characteristics at 270 K were compared to those of a QD laser whose structure is nominally the same as that of the TI QD laser except that it lacks the TI region. $I - V$, $P - V$ and $P - I$ characteristics of the LDs were measured in the temperature range of 210 to 320 K, using an Agilent 4155 C Semiconductor parameter analyzer. The optical power of devices was measured with a calibrated photodetector.

The experimental data enabled to extract several important electrical parameters of the TI QD diode. These parameters were used to simulate the $I - V$, $P - V$ and $P - I$ curves using a detailed model the results of which confirm the measured results. No complete desceription exists in the literature addressing the influence of the tunnelling process on the static characteristics and parameters of TI QD lasers. Detailed studies of the $I - V$, $P - V$ and $P - I$ characteristics are shown here to be effective tools that when accompanied by a model to discriminate among carrier transport mechanisms of which tunneling is but one process. This is also correlated with the emission properties of the TI QD laser to present a comprehensive description of the device properties.

## II. Device structure and experimental procedure

The TI QD laser comprised six layers of highly uniform $InAs$ QDs with a density of $3 \times 10^{10} cm^{-2}$. Each 2 nm thick QD layer is accompanied by a 3 nm wide QW separated from the QDs by a 2 nm thick $InAlGaAs$ barrier layer [13]. The energy band diagram of the TI QD laser we studied is shown in Fig. 1(a). An atomic force micrograph of a single QD layer is depicted in Fig. 1 (b) and Fig. 1 (c) and d) are high-resolution transmission electron microscope cross sections.

The laser cavity was formed by a 2 $\mu m$ wide, 330 $\mu m$ long ridge waveguide. Reference lasers with nominally identical QDs but with no TI layers were fabricated for comparison. $I - V$, $P - V$ and $P - I$ were measured using standard techniques and equipment.

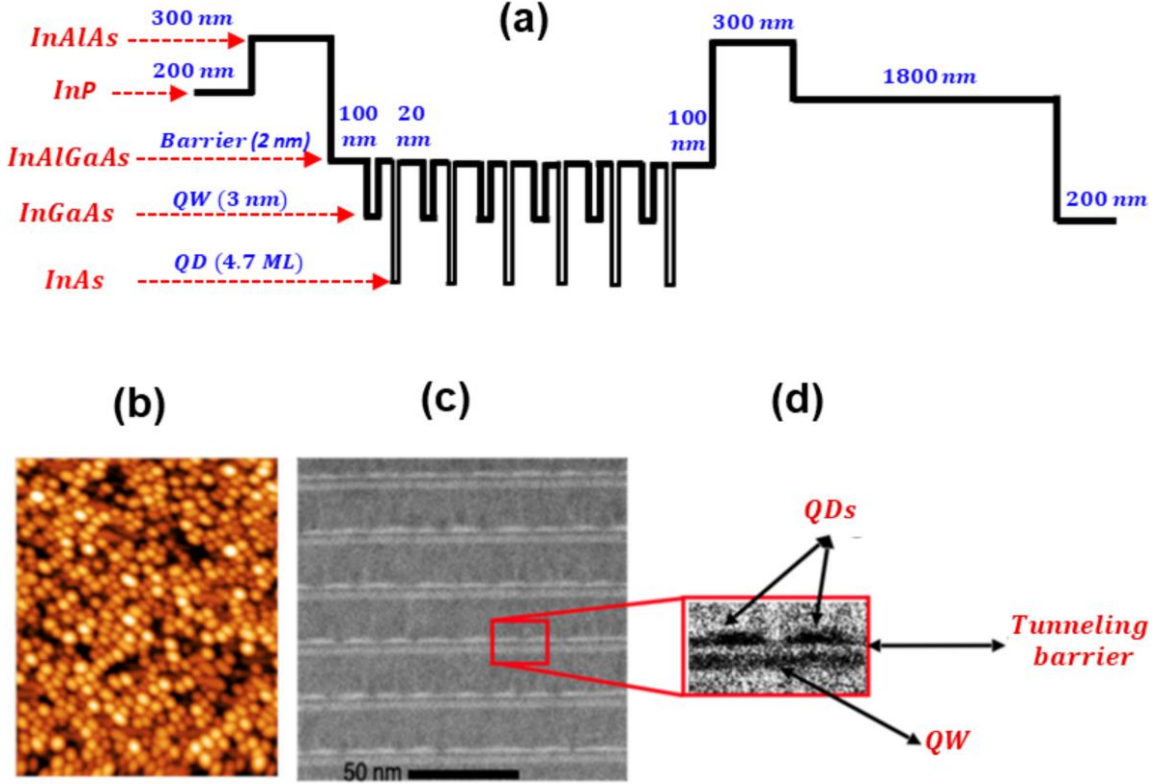

Fig. 1 Device structure. (a) Energy band diagram. (b) Atomic force microscope image of a single QD layer. Site specific cross section measured by high resolution transmission electron microscopy of the entire stack (c) and of the QDs and the TI region (d).

## III. Experimental results and discussion

### 1. Current voltage characteristics

$I - V$ and the corresponding $\alpha - V$ characteristics, measured in the temperature range of 210 K to 320 K are shown in Fig. 2. Five regions are clearly distinguishable in the $\alpha - V$ curves. In the first three, the semi-logarithmic $I - V$ curves are nearly linear but with different slopes while in the fourth they exhibit a highly nonlinear behavior. These translate to three maxima (or bends) in the $\alpha - V$ characteristics which vary with increasing temperature. The third maximum represents a change from an exponential to a power-law and / or a linear dependence of the current on voltage. This region is followed by a monotonic decrease of $\alpha$ leading to a sharp discontinuity at a temperature below 290 K. The height of the kink at the discontinuity decreases when the temperature increase. The height of the first peak is highly temperature sensitive. This dependence is weaker for the second peak while the third peak height is basically temperature independent. The voltages at which the peaks occur shift towards lower voltages as the temperature rises with the shift of the third peak being the largest. The significance of the various peaks is discussed later in the paper.

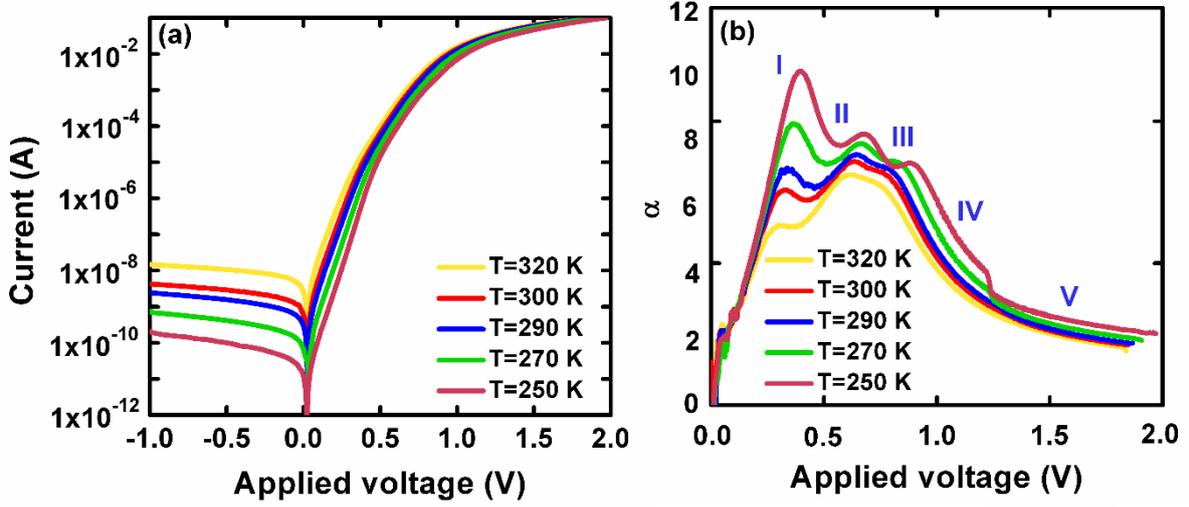

Fig. 2. Measured temperature dependent $I - V$ (a) and the corresponding $\alpha - V$ characteristics (b) of a TI QD laser. The numbering in (b) indicates areas corresponding to various current flow mechanisms.

The first region where the bias is smaller than 0.5 V, including the reverse bias part of the $I - V$ curve (Fig. 2 (a)), exhibits the properties of a conventional diode including the saturated and exponential branches which are strongly temperature dependent. In this range, the $I - V$ curves follow the classical Shockley-Read-Hall (SRH) model [14, 15]

$$I = I_S exp \left( \frac{qV}{E_T} \right) \quad (1)$$

$I$ and $I_S$ are total and reverse saturation current densities. $E_T = nkT$ denotes the characteristics energy (or the transparency energy of the barriers) in an ideal diode, where $n$ describes the predominant conduction process at the $p - n$ junction, which in accordance with the SRH model varies between 1 and 2. The parameter $n$ depends on the dominant current flow mechanisms; $n = 1$ for carrier diffusion and $n = 2$ for the generation-recombination processes of electrons and holes via deep energy levels within the energy gap ($E_g$). q, $V$, $k$ and $T$ are the elementary charge, applied voltage, Boltzmann constant and absolute temperature, respectively. When $n$ is larger than 2, the dominant carrier transport mechanism is associated with tunneling where the current has but a minor dependence on temperature. A simplified formalism, suggested in [16], has a similar form as equation (1) but with a distinct pre exponential saturation current value describing the diode current when limited by tunnelling [17, 18]

$$I_T = I_{ST} exp \left(\frac{qV}{E_T}\right), \qquad (2)$$

Where the tunneling saturation current is

$$I_{ST} = I_{0ST} exp \left(\frac{-qV_{bi}}{E_T}\right) \qquad (3)$$

with $V_{bi}$ representing the built-in voltage and and $I_{0ST}$ is a temperature and voltage insensitive constant.

The SRH and tunneling models are only valid at low voltages and current densities, where the dependence of the current on the voltage is exponential. For high voltages and currents, the dependences deviate from an exponential form (as in regions 4 and 5 of Fig. 2). These regions are most important since they include the bias levels where first spontaneous and then stimulated emission occur. In an analysis based on the $I - V$ characteristics, the linearity amounts to a voltage drop across a series resistance caused by the bulk semiconductor and metal-semiconductor junction [19-22]. However, in the general case, since the gain region of laser diodes is within of intrinsic ($I$) area of the $P - I - N$ structure, a high injection level of carriers and hence non-linearity is expected [23].

The nonlinear contribution to the $I - V$ characteristic can be approximated by a power exponent law [23, 24]

$$I = B_i \left(V_{tot,i} - V_{ci}\right)^m \quad (4 \text{ a})$$

Where

$$V_{ci} = (E_g - E_t) + \alpha_c T \quad (4 \text{ b})$$

is the forward voltage drop at both junctions (contact potential) [23, 25, 26] of the $P - I - N$ diode; $B_i$ is a coefficient for which the dependence in (4 a) sets in. $E_t$ and $\alpha_c$ are respectively, the impurity ionization energy and temperature coefficient that characterized the variation of the contact potential. Equation (4 a) approximates the recombination current in a $P - I - N$ diode. $m$ values of 2 or 4 relate respectively to devices with short or long compensated (intrinsic) layer relative to the carrier diffusion length [23, 24] The approximation (4) is the same as the one used to characterize current flow in light emitting diodes with $P^+ - I - N$ or $P - I - N^+$ structures having either narrow or wide compensated regions [26]. This is related to monomolecular or bimolecular recombination in the compensated and / or in the highly doped junctions which also yield power exponent values of $m = 2$ or $m = 4$, respectively [25-27].

The total applied voltage ($V_{tot,i}$) is formulated as [12, 23-26, 28]

$$V_{tot,i} = \frac{nkT}{q} \ln \left( \frac{I}{I_S \ or \ I_{ST}} \right) + A_i I^{1/m} + I R_S + V_{ci} \quad (5)$$

With $A_i = 1 / B_i^{1/m}$. The first term in (5) is the voltage drop across the junctions adjacent to optical confinement layers (OCL). The second and third terms describe potential drops across the intrinsic layer due to a nonlinear current flow, and the linear series resistance while the fourth originates in the contact regions. Under practical operating conditions, diode lasers do not reach the linear regime, therefore, the $I R_S$ term in (5) is insignificant, especially at temperatures above 300 K though it may have some impact at low temperatures. Nevertheless, the procedure for extracting the diode parameters given below considers the effect of a series resistance, $R_s$.

The general formalism for $\alpha - V$ in the presence of a series resistor is:

$$\alpha = \frac{qV}{E_T} \frac{1}{(1 + \frac{qIR_S}{E_T})} \quad (6)$$

In order to extract parameters in the moderate applied voltage regimes, where only the exponential part and the effect of the series resistance play a role in the $I - V$ characteristic, it is common to use the maximum value of the $\alpha - V$ characteristic ($\alpha_m$) by employing the formalism [28]:

$$\frac{E_T}{q} = V_m \frac{(\alpha_m - 1)}{\alpha_m^2}; \ R_S = \frac{V_m}{I_m} \frac{1}{\alpha_m^2}; \ I_S \ (or \ I_{ST}) = I_m \exp\left(-(1 + \alpha_m)\right) \quad (7)$$

where, the values of $I_m$ and $V_m$ correspond to the $\alpha_m$ obtained from the function (6).

However, in the present case of multi-exponential contributions, in particular at the high temperatures where the peaks in the $\alpha - V$ curve are broad so that their maximum is ill-defined, it is hard to extract reliable, constant parameters. These parameters can be obtained however, from the values of $I, V$ and $\alpha$ corresponding to the non- monotonic regions, by using:

$$\frac{E_T}{q} = V \frac{(\alpha + \theta - 1)}{\alpha^2}; \;\; R_S = \frac{V}{I} \frac{1 - \theta}{\alpha^2}; \;\; I_S \text{ (or } I_{ST}) = I \exp\left(-\left(1 + \alpha \frac{(\alpha - 1)}{(\alpha + \theta - 1)}\right)\right) \quad (8)$$

where the $\theta = \frac{V}{\alpha} \frac{d\alpha}{dV}$, is the slope of the $\alpha - V$ characteristics in a linear scale. From these, one finds the $E_T$, $R_S$ and $I_S$ or $I_{ST}$ values which are constant at a given temperature. These parameters are then used to simulate the theoretical models expressed by equations (1), (2), (3) and (5).

An analysis at bias levels where the exponential terms are no longer valid, namely the regimes of high carrier injection, where first spontaneous and then stimulation emission start (before and after the discontinuity observed in the $\alpha - V$ curves (Fig. 2 (b)), requires knowledge of the parameters $m$, $V_{ci}$ and $A_i$ in equation (4). Assuming a voltage independent $m$ ($\frac{dm}{dV} = 0$) the three parameters can be found using the procedure given in [12]:

$$m = -\frac{\alpha}{(\theta - 1)}; \;\; V_{ci} = \frac{\theta V}{(\theta - 1)}; \;\; A_i = \frac{1}{I^{1/m}}\left(-\frac{V}{(\theta - 1)}\right) \quad (9)$$

The second term in eq. (5) represents the nonlinear contribution to the voltage drop $V_I = V_{tot,i} - V_{ci}$ (see equation (4)), which falls on the intrinsic layer of the diode [23, 24]. $V_I$ is equivalent to $V_{nl} = IR_{nl}$, introduced in [29] as a way to take into account the injection induced conductivity. Therefore, in the high injection mode described by equation (4 a), the nonlinear resistance $R_{nl}$ adds to the linear series resistance $R_s$. Differentiation of $V_I$ (or $V_{nl}$) with respect to current yields $R_{nl}$, which is coupled to the second, non-linear, term in (5) through the relation

$$R_{nl} = \frac{A_i}{m} I^{\left(\frac{1-m}{m}\right)} \quad (10)$$

The value of $R_{nl} = \frac{d}{S} \frac{1}{q \mu_e N (1 + \frac{\mu_h}{\mu_e})}$ depends on the parameters of the intrinsic layer: injected carrier density ($N$), electrons and holes mobilities ($\mu_e$ and $\mu_h$, respectively), the area $S$ and the thickness $d$ of the waveguide region. Equation (10) allows to derive the important relation between injected carrier density and current in the radiative emission region and hence to establish the dominant recombination mechanism

$$N = \frac{Cm}{A_i} I^{\frac{(m-1)}{m}}, \qquad (11)$$

Where,

$$C = \frac{d}{S} \frac{1}{q\mu_e(1+\frac{\mu_h}{\mu_e})} \qquad (12)$$

The relation between carrier density and current, as in equation (11) has been used before, for example in [30] but without considering the high injection regime which is described in (12).

The parameters extracted from equations (7-9) are used to simulate the exponential, nonlinear, and linear parts of the measured $I - V$ characteristics. One approach to analyze the non-monotonic behavior of the $a - V$ characteristics is to model the current as flowing through a cascade of nonlinear elements, each associated with the structural and physical parameters of the device (layer geometry, interlayer barriers, properties of the quantum dots, doping levels, etc.).

A simple electrical model is shown in Fig. 3. The input current in the second section is assumed to be limited at low applied voltages by the largest differential resistance $R_{j1}$ of the first section. It is divided into two parallel components (through the equivalent resistance $R_{j2}$). One part is the leakage current associated with thermally activated carriers over the potential barrier into the region of quantum dots predicted in [30] and the second parallel branch is associated with the tunnelling process, the occurrence of which is possible only under certain conditions [3-5]. The details of the mechanisms of these two current components are discussed below when analyzing the experimental and $a - V$ characteristics. Fig.4 shows $I - V$ characteristic and $a - V$ characteristics and the corresponding results of the approximation for a temperature of 270 K. The three distinct exponents making up the overall characteristics are clearly seen. Deviations of the curves from linearity (shown in dash lines in Fig. 3) result from the effective series differential resistance connected to the exponential parts.

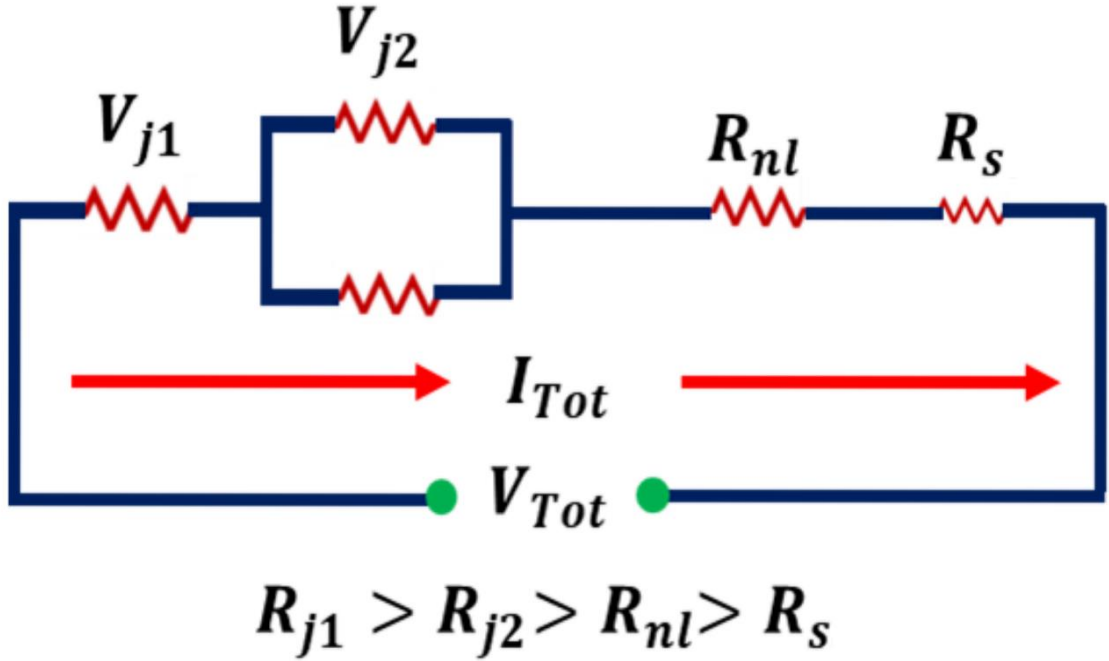

Fig. 3. Schematic equivalent circuit of the TI QD laser. $R_{j1}, R_{j2}$ are characteristic differential resistances of the junctions arising from different current flow mechanisms at different ranges of the applied voltage.

The excellent fits in all three regions testifies to the validity of the exponential current increase with voltage described by equations (1) or (2) and (3), independent of temperature.

The inserts in Fig. 4 (a) and (b) show once more the measured $I - V$ and $a - V$ characteristics, also for 270 K, together with the simulated results that do not include the effective series differential resistance. The three exponentials represented in semi logarithmic plots are fit by linear curves with different slopes. The linear curve fits the first region in the first 0.5 V while in the second and third regions, the fits cover smaller voltage ranges, 0.2 V and 0.3 V, respectively. This testifies clearly to the need to include the series resistance as in Fig. 4 (a) and (b).

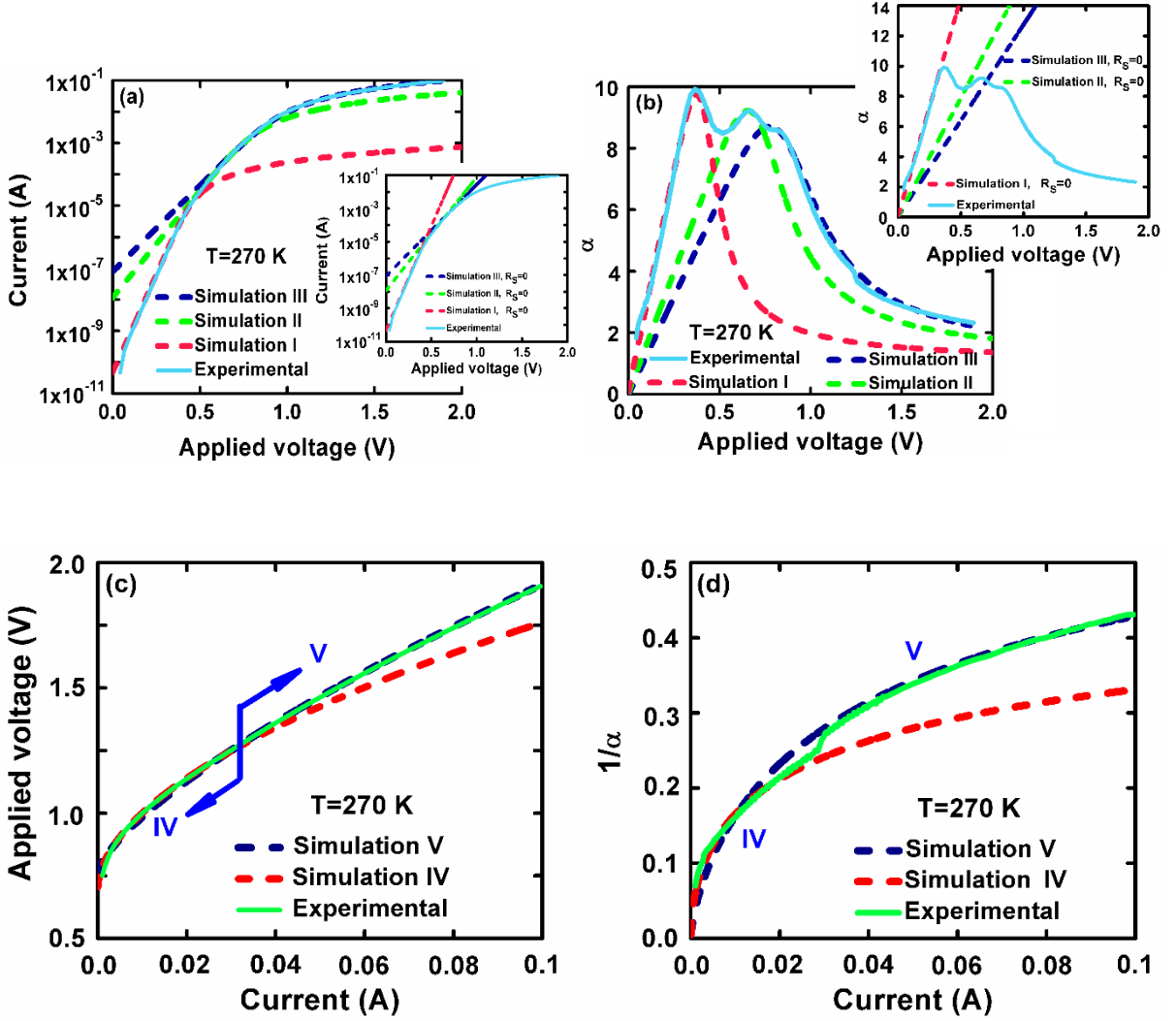

Fig. 4. Measured, at 270 K, (a) and calculated $I - V$ and the corresponding $\alpha - V$ (b) characteristics. The experimental curves are solid lines and the calculated ones are shown in dash lines. The calculation considers the exponential parts of curves and the influence of the series resistance, while the insets describe simulation based on exponential dependencies only with $R_s = 0$. Experimental (c) $V - I$ and (d) $1/\alpha - I$ dependencies with simulated curves for the high injection regimes in accordance with [36, 37] and [22] (equations (13) and (4 a).

Fig. 4 (c) shows a portion of the $V - I$ characteristics, measured at 270 K, for voltages beyond the experimental part of the $V - I$ characteristics where the current is in the range of 0.01 mA to 0.1 mA. Fig. 4 (d) describes the corresponding $1/\alpha - I$ curves (following the third maximum in Fig. 2(b)), presented in linear coordinates. The $1/\alpha - I$ dependencies are similar to the current derivative $I \frac{dV}{dI} - I$ formalism described in [21, 31], which is used to extract the threshold current from the $V - I$ rather than the $P - I$ characteristics. The $1/\alpha - I$ curves (Fig. 4 (d)) increase monotonically in contrast with the $\alpha - V$ reduction beyond the third peak. However, the rate of rise (or reduction) is different below and above threshold. The threshold appears as a clear vertical rise (or reduction) at an applied bias of about of 1.27 V (at 270 K). It remains almost constant at 1.25 V in the temperature range of 280 K to 320 K as seen in Fig. 2 (b). The post exponential parts of the experimental curve approximated in accordance with equation (4) using $m$, $V_{ci}$, and $A_i$ values extracted by equations (9) are shown in solid lines. The simulated results fit the experimental curves very well.

Determination of the current flow mechanisms in all regions is based on an analysis of the temperature dependencies of the saturation current and the characteristic energy (or ideality factor) determined by equations (7) and (8), as shown in Fig.5. The first region exhibits an increase in characteristic energy $E_{T1}$, extracted from the first $I - V$ branches, with an exponent of the order of 1.07 (see Fig. 5(a) which is very close to the predicted linear dependence of $E_T$ on temperature in the SHR current flow mechanism. In this regime the ideality factor is roughly 1.465 and in the range of 210 K to 320 K, it is obtained independent on temperature (see Fig. 5 (b)). The maximum deviation of an ideality factor from the mean value is $\pm 2\%$. This indicates a mix of thermally activated diffusion and a generation-recombination dominant contribution. The characteristic energy $E_{T2}$ (related to the second exponential branch of the I-V characteristic) changes, unlike $E_{T1}$, sublinearly with temperature, with a slope of 0.2. This is due to temperature sensitivity of the ideality factor, which reduces from 3.35 to 2.428, when the temperature rises from 210 K to 320 K. The third exponential curve is characterized by a large $E_{T3}$ and ideality factor, both of which are strong nonlinear functions of temperature. The deviation of characteristic energies from linear increases with temperature is associated with the behavior of the temperature dependence of the ideality factor which means that the SHR-classical model is replaced by the tunneling mechanism [16, 17, 19]. Note, that the value of $n_2$ approaching 2 at elevated temperatures denotes that in the second exponential regime the recombination and tunneling processes are not strongly limited as that in the third exponential region.

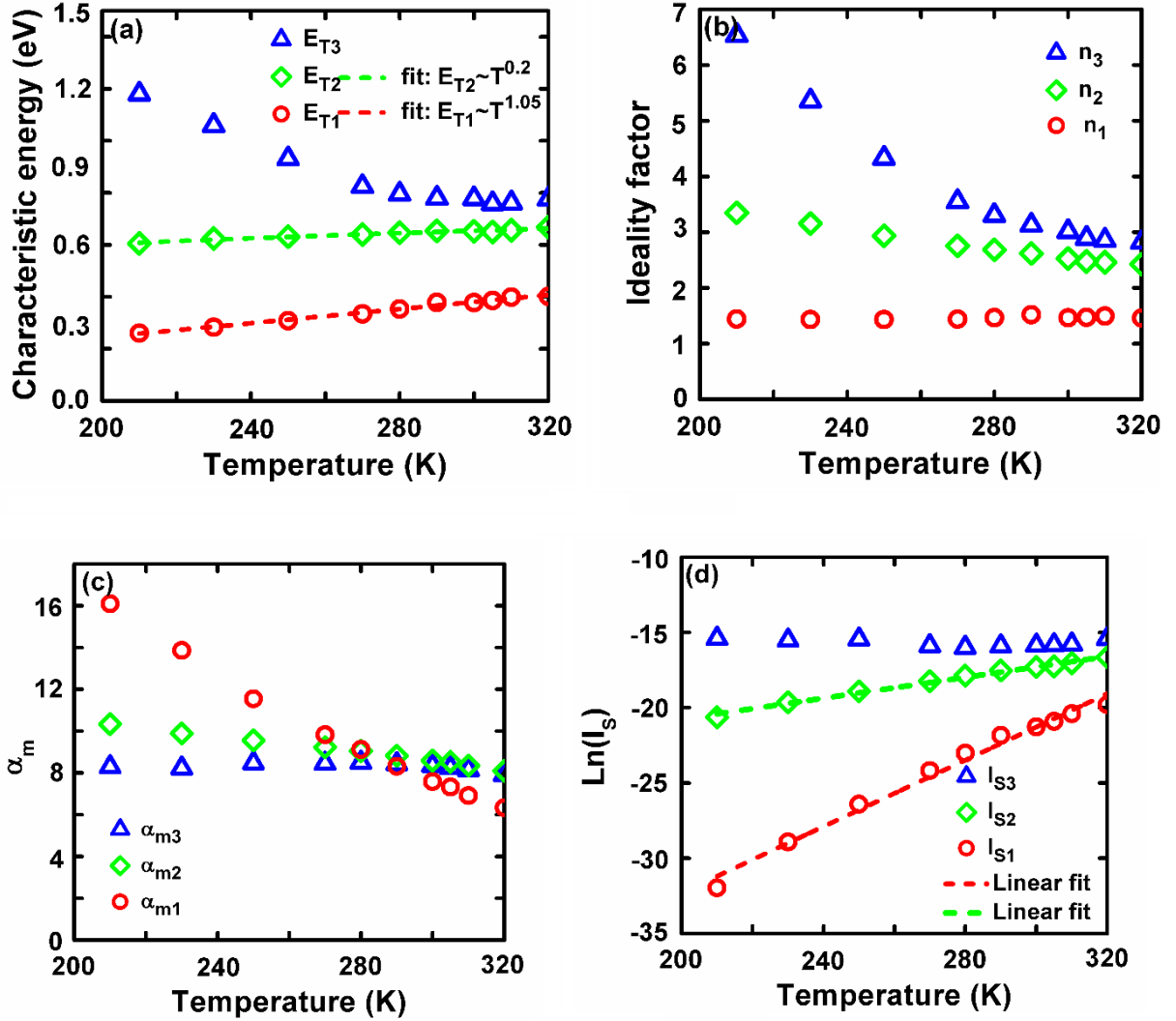

Fig. 5. Temperature dependence of the extracted characteristic energy (a), ideality factor (b), maximum slope (c), and saturation current (d), corresponding to different exponential parts of the $I - V$ characteristic (see numbering in Fig. 2(b)).

An additional feature is the weaker temperature dependence of the maxima, or bends, in Fig. 2 related to the second and especially to the third branches of the characteristics, in sharp contrast to the temperature dependence of the slope in the first exponential region, where the maximum changes from 16 at 210 K to 6 at 320 K (see Fig. 2 (b) and Fig.5 (c)). This observation discriminates the standard SHR diode model from carrier tunneling through the barrier.

Finally, in the considered range of the applied voltage, the logarithmic value of the saturation current, extracted from the first exponential region of the measured forward I – V characteristic (using equation (8)), varies linearly with temperature, and yields a band gap ($E_g$) of about 1.1 eV. This value was estimated using the equation $I_s \sim T^{3+\frac{b}{2}} exp\left(-\frac{E_g}{nkT}\right)$ [32, 33] which holds for the classical SHR model. $b$ is an integer, the variation of which with temperature is determined by the dependence on temperature of the charge carriers mobility. The extracted band gap is slightly larger than that determined from the measured reverse I – V characteristics (see Fig. 2 (a)) which is 1.018 eV, but is close to the theoretical band gap of the $InAlGaAs$ barrier layer which is 1.07 eV [34]. The value used for $b$ was -3.

The temperature dependence of the saturation current $I_S$, extracted from the second exponential is linear, as in the first region, but the slope is smaller. In the third region, $I_S$ is practically independent of temperature. A weak temperature dependent characteristic energy in the second and third regions and therefore an ideality factors which is larger than 2, (see respectively Fig. 5 (a) and Fig. 5 (b)), together with the exponential bias dependent relationship are distinctive features of trap assisted tunnelling in $P-N$ or $P-I-N$ light emitting diodes [17, 19, 35].

Fig. 6 presents the dependence of the forward voltage on temperature for various current levels. Each curve represents a current of a flow mechanism which is identified in the experimental I – V characteristics (see Fig. 2). According to the SHR model, the voltage drop across the junction (first term of equation (5)), depends linearly on the ambient temperature [33]. Extrapolation of the linear curves to 0 K yields therefore the band gap energy of the junction material, which determines in turn, the nature of the $I-V$ characteristics. Fig. 6 shows that only the first section in the $I-V$ curve, is linear over a wide temperature range, from 170 K to 320 K. The second curve deviates slightly from linearity below 250 K, and the third is highly nonlinear even below 270 K. This confirms once more that the SHR model is only valid, over a wide temperature range, for the low bias levels.

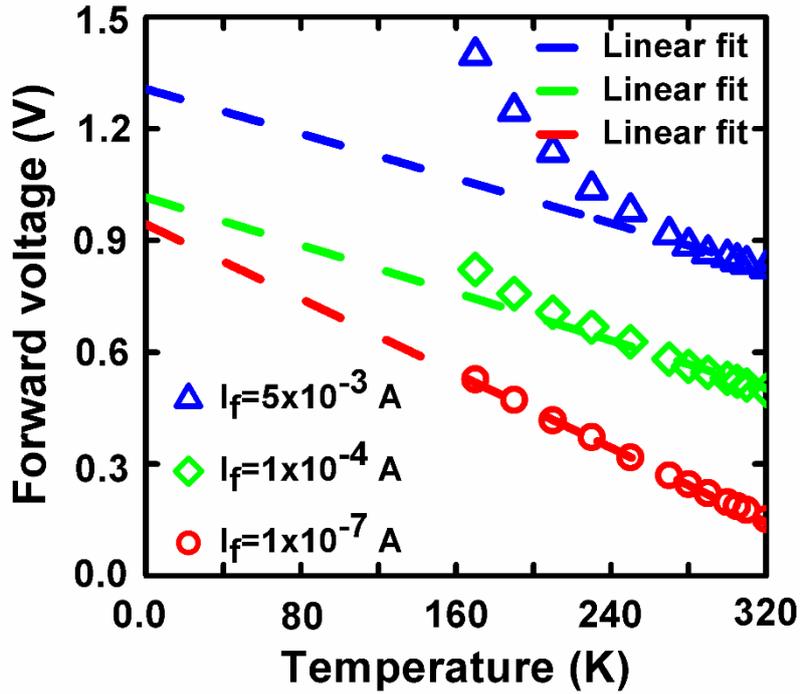

Fig. 6. Temperature dependence of the experimental forward voltage obtained by varying the injection currents corresponding to the different exponential sections of the measured $I - V$ characteristics.

Extrapolating the first part the I-V characteristic to T=0 K yields a bandgap of 0.975 eV, close to the band gap of the $InAlGaAs$ barrier layer [34] and to that obtained from measured saturation current. A reasonably close value of 1.016 eV is obtained by extrapolation the second, partially fitted, curve (in the temperature range from 270 K to 320 K). The third region is very different, even its linear part, expanded between 280 and 320 K yields a large bandgap of about 1.1595 eV. The discrepancy may be due to the placement of the tunneling injection layer between the barrier and the quantum dots layers, which facilitates carrier penetration via tunneling into the QD region. Another reason may be the influence of the QD regions which are enclosed between high potential barrier layers [36].

The observed behavior of the barrier transparency energy (introduced as a general parameter that determines the tunneling process), that decreases with temperature cannot be explained by an excess current due to native or exited defect trap states in the device. Such a parallel conduction path requires a temperature dependence of the characteristic energy which is opposite to the one observed. Also, as a rule, excess current branches appear on the $I - V$ characteristic before the thermally activated diffusion and / or generation-recombination currents. Moreover, current flow through parallel channels, and their effect on the $I - V$ characteristic, disappears at high voltage and temperature, when the exponential branches associated with intrinsic potential barriers at the metal-semiconductor, $P - N$ or $P - I - N$ interfaces are more sensitive to temperature and applied voltage variations. In contrast, the measured $I - V$ characteristics do not include any parallel components of a linear or tunneling nature but start with an exponential branch limited only by a mixture of diffusion and recombination currents. Further, a parallel excess current implies a symmetrical occurrence for reverse bias. Fig. 2 (a), exhibits an asymmetric nature with respect to the reverse current branch, indicating once more the difference in nature from other light-emitting devices [7]. Limitation of carrier injection into the active region from the side of the barrier part of the device (resulting from the SHR mechanism), leading to the first exponential region of the $I - V$ characteristic, requires a certain applied voltage to initiate two separate TI processes which take place simultaneously: resonant tunnelling [5], which make use of a hybrid state between the lowest QW energy state and an excited state of the QDs, followed by fast relaxation to the QD ground state, and non-resonant tunnelling where phonons mediate direct tunnelling of QW carriers to the QD ground state. These two tunnelling injection mechanisms, are responsible for the second and third exponential branches. These mechanisms differ from the process of defects (traps) assisted tunnelling, which cause parallel leakage currents and are the reason for the observed type of dependence of the characteristic energy on temperature.

The post-exponential sections of the experimental $I - V$ characteristic in the current range from 0.0023 A to 0.1 A, and the corresponding $a - V$ curves (beyond the third maximum), measured at 250 K, shown in Fig. 4 (a) and (b)), were approximated by equation (4) using the parameters $m, A_i$ and $V_{ci}$, extracted from the experimental data by equation (9). The observed discontinuity results from pinning of the modal gain at threshold. Pinning is never complete due to the gain nonlinearity [37], but for the present analysis, the assumption of gain pinning is acceptable. As the current increases above the threshold, stimulated emission pins the electron and hole quasi Fermi level separation and hence, the voltage across the diode is essentially constant [12, 20, 21, 29, 31]. The temperature dependence of the extracted $V_{ci}$ (see Fig. 7) is linear and when extrapolated in accordance with equation (4 b) to $T = 0 K$ yields 1.01 $eV$, close to band gap of $InAlGaAs$ barrier layer [12, 34]. The same value was obtained from the measured saturation current or forward voltage versus temperature. The power exponent $m$ dependencies on temperature for the sections related to spontaneous and stimulated modes are also shown in Fig. 7, demonstrates linear reduction with temperature. Finally, the value of $B_i$ is practically independence on temperature. This is due to the opposite temperature dependencies of the parameters that define its value [26].

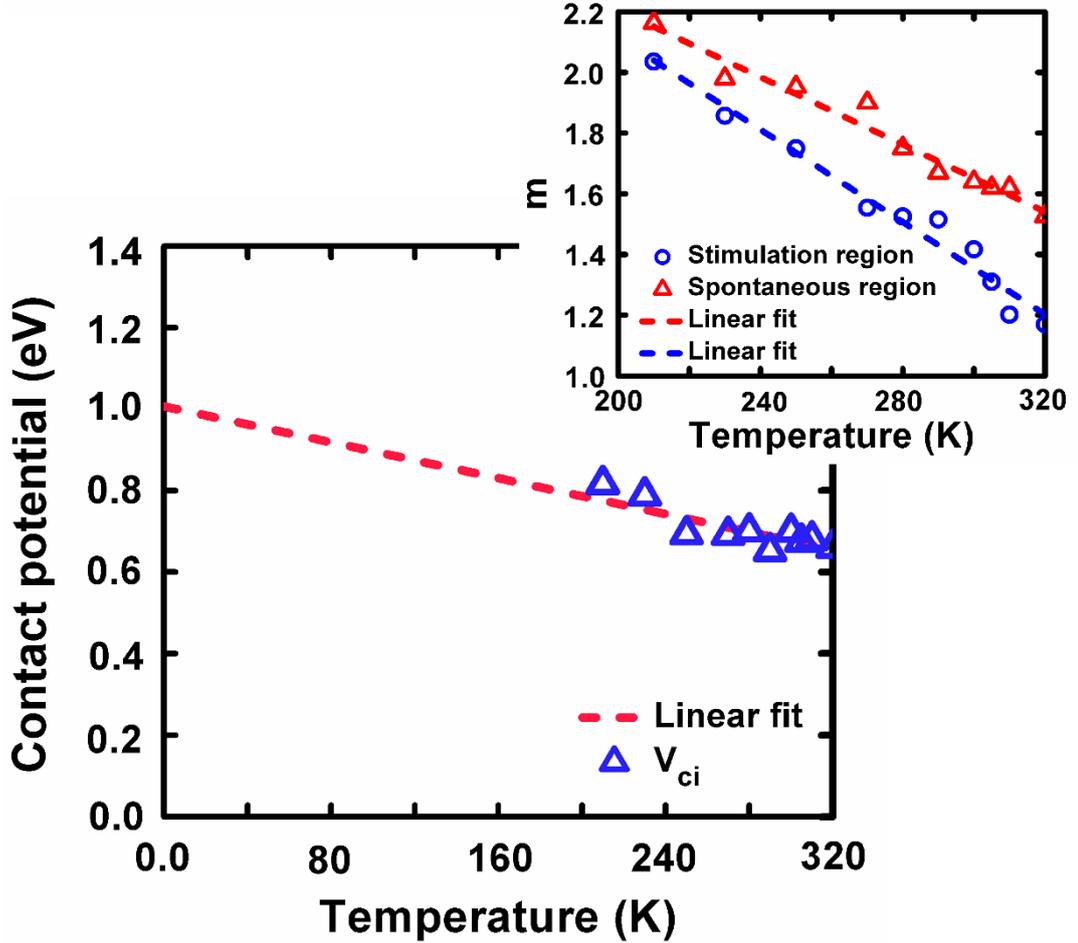

Fig. 7. Temperature dependence of the voltage drop on junctions and power exponent $m$ in the high injection regime.

In order to clarify the distinctive charge transport features of the TI QD laser, we compared its characteristics to those of a nominally identical QD lasers with no TI. The $I-V$ and $a-V$ characteristics are shown in Fig. 8 (a) and (b). There is one marked difference, the number of linear regions in $Ln(I)-V$ and $a-V$ characteristics. The QD laser reveals only two maxima. for QD LD corresponding to variation of current flow laws in exponential regions. Beyond the last maximum, the curves are similar and include two parts separated by jump-like reduction (see discontinuities in the $a-V$ curves in Fig. 2 (b)).

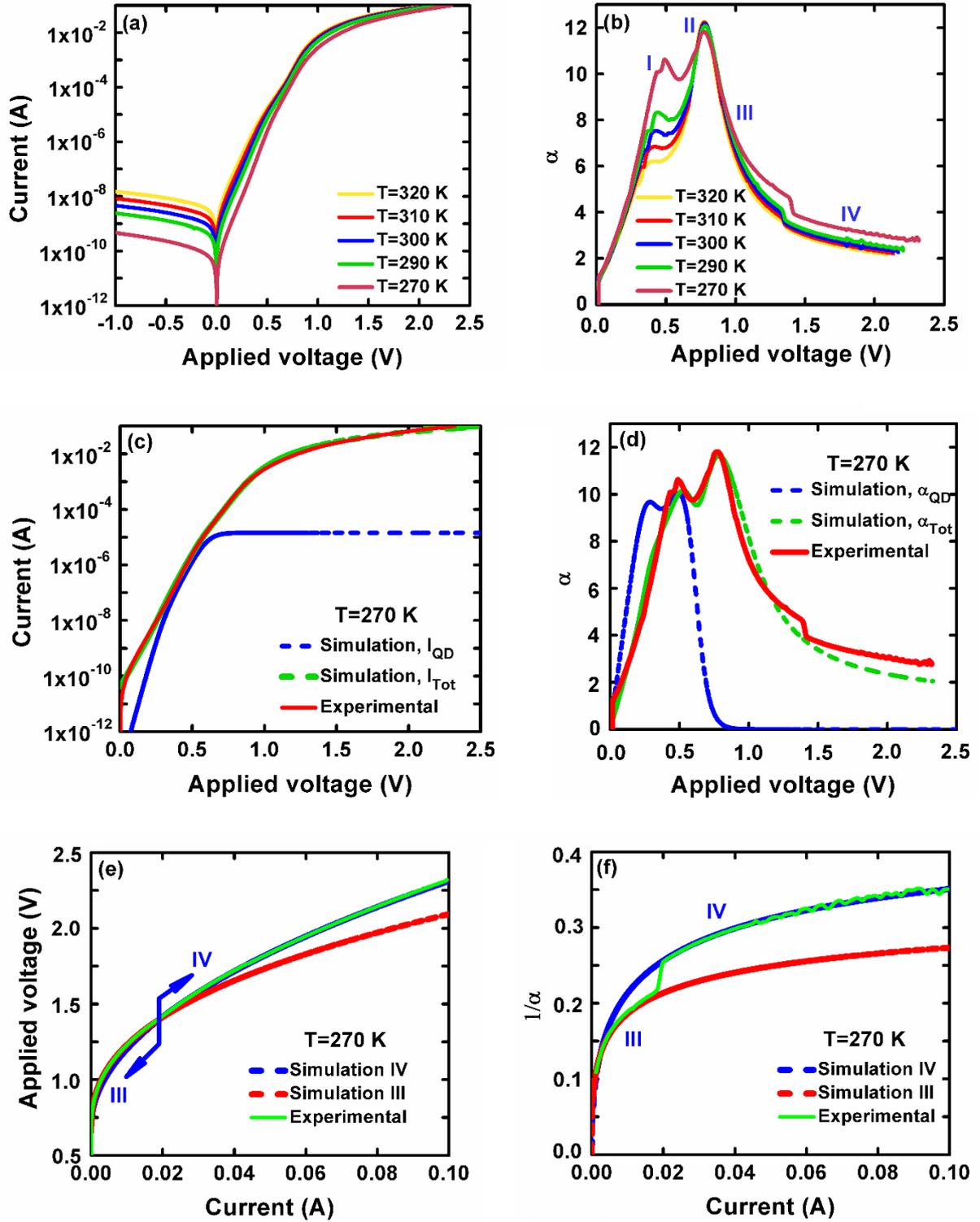

Fig. 8. Measured $I - V$ (a) and the corresponding $\alpha - V$ (b) characteristics of a QD laser (with no tunneling region), measured at different temperatures. (c) Simulated $I - V$ and (d) $\alpha - V$ dependencies, defined by the exponential character of the parallel branches of currents flowing through the optical confinement layer and the QD regions considering the series resistance. Experimental (e) $V - I$ and (f) $\frac{1}{\alpha} - I$ dependencies with the corresponding simulated curves

in the high injection regimes in accordance with [30, 38] and [23] (equations (13) and (4 a)). The numbering in (b) indicates areas corresponding to various current flow mechanisms.

Current flow in a QD laser was modeled by Asryan and Suris [30, 38], which describe two parallel components which depend exponentially on the applied voltage:

$$I_{tot} = \frac{qN_S}{\tau_{QD}} f_n f_p + I_s \exp\left(\frac{qV_{tot}}{kT} - 1\right), \quad (13)$$

Where $f_n$ and $f_p$ are respectively the mean electron and hole occupancies in the QDs. $N_S$ is the QD density in a single layer and $\tau_{QD}$ is the radiative lifetime in the QDs.

The first term in (13) is the current flow through the QDs at low voltages and the second one represents current through the optical confinement layer assumed to be a diffusion current, where the ideality factor is 1, in accordance with the SRH model. In [12], we have proven experimentally the required addition of the high injection non-linear regime [23, 26] in the region where radiative recombination takes place (see equation (4)).

A simulation of the measured results at 270 K that uses (13) in combination with (5) and (8) (not including the high injection regime (4)), is shown in Fig. 8 (c) and (d) and demonstrates the validity of both [31, 38] and [23] models in the appropriate bias ranges. The main difference between the diodes is the series SRH branch with a clear temperature activated character, a constant ideality factor and a strong tunneling contribution in the TI QD lasers for bias levels prior to radiative recombination. The post-exponential part related to the high injection mode was approximated by equation (4), as for the TI QD LD. The results of modeling the experimental $V - I$ characteristics and the corresponding $1/\alpha - I$ curves (after the second maximum in Fig. 8 (b)) shown in Fig. 8 (e) and (f) exhibit the same behavior as in the TI QD diode (see Fig. 4 (c) and (d)).

The temperature dependence of the characteristic energy of the convention QD laser, and hence the ideality factor, extracted from the parameters of the second regions in the $I - V$ and $a - V$ characteristics (see Fig. 8), using equations (7) and (8), is shown in Fig. 9.

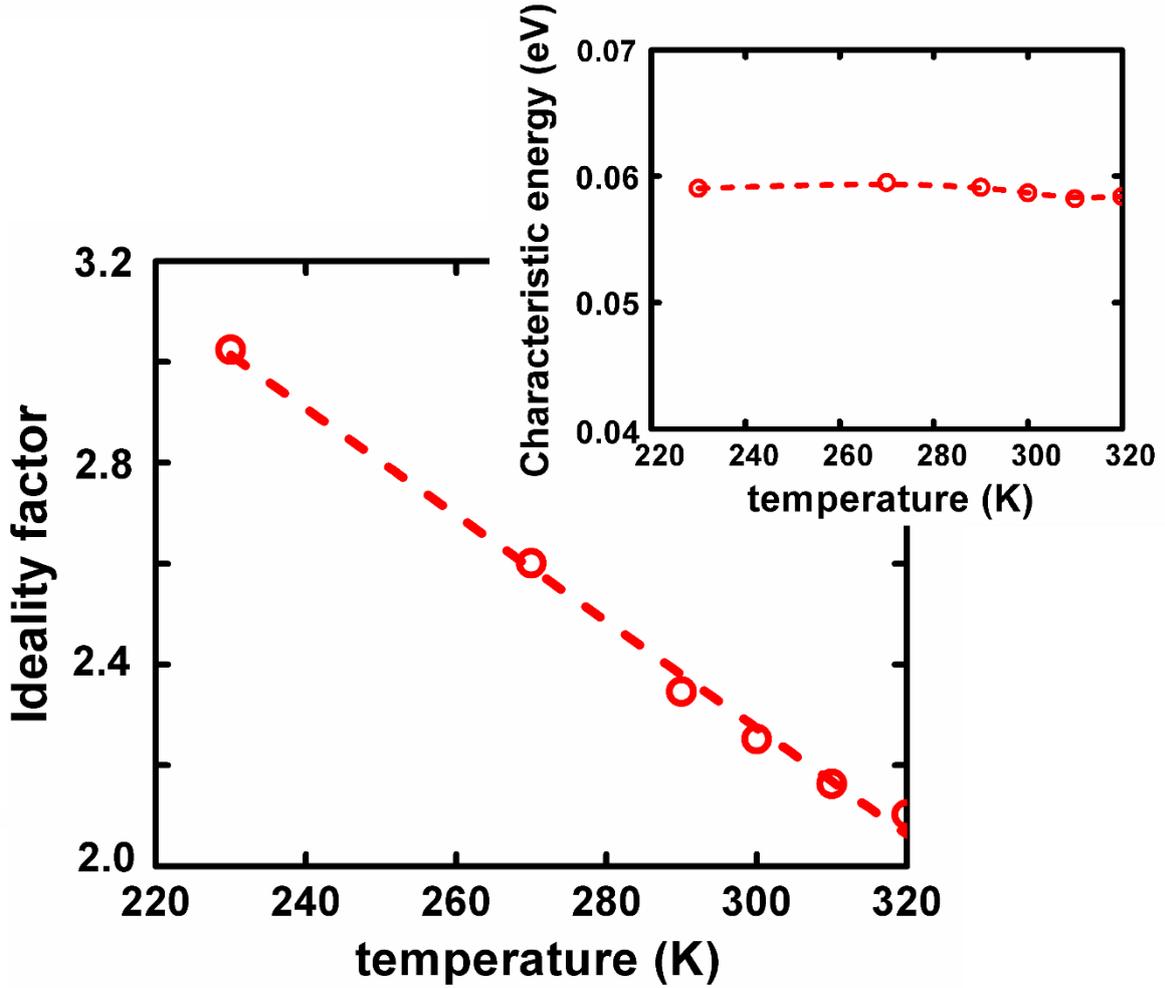

Fig. 9. Temperature dependence of the extracted characteristic energy and ideality factor corresponding to the second parts of the $I - V$ characteristic related to the optical confinement layer of the QD laser.

The curves show that the slight change in $E_t$ and the increase of the ideality factor from 2.1 at 320 K to 3 at 230 K is identical to that observed for TI QD laser in the second region of applied voltage (see Fig. 5 (a) and (b)). Since this region in the I-V and α-V curves of the QD laser (see Fig. 8) follows a parallel current path caused by the QD region, we postulate that the second region in the TI QD laser also results from a mixed effect of thermally activated current flowing through the QDs (see Fig. 8, first branch) and the optical confinement region, described in [12, 30]. This current is added to the parallel tunneling component.

### 2. Optical output power voltage characteristics

The $I - V$ data of laser diodes are directly related to optical output power versus current and applied voltage dependencies the $P - I$ and $P - V$, respectively. The most common presentation of the electro optic characteristics of a diode laser is the $P-I$ measurement. Alternatively, the $P-V$ characteristics, that are not commonly used and for which there is no reported analysis, correlate directly to the I-V measurements thereby shedding light on intricate details of the radiative recombination that are not observable in $P-I$ curves.

$P-V$ and $\gamma-V$ characteristics of the TI QD laser, measured at 230 K to 310 K are shown in Fig. 10. Three distinctive regions are observed; The first and third regions are separated by an abrupt increase in output power and a sharp peak in $\gamma$ which represents of course the oscillation threshold. The threshold voltage corresponds to the discontinuity in the $I-V$ curve. $\gamma$ drops fast on both sides of the peak. Lower voltages represent the spontaneous emission region and high voltages correspond to the high injection, high output power regime. The entire voltage region where any kind of radiative recombination takes place is beyond the exponential part of the $I-V$ curve. Practically both disappear at 320 K which means that the laser experiences a very soft threshold.

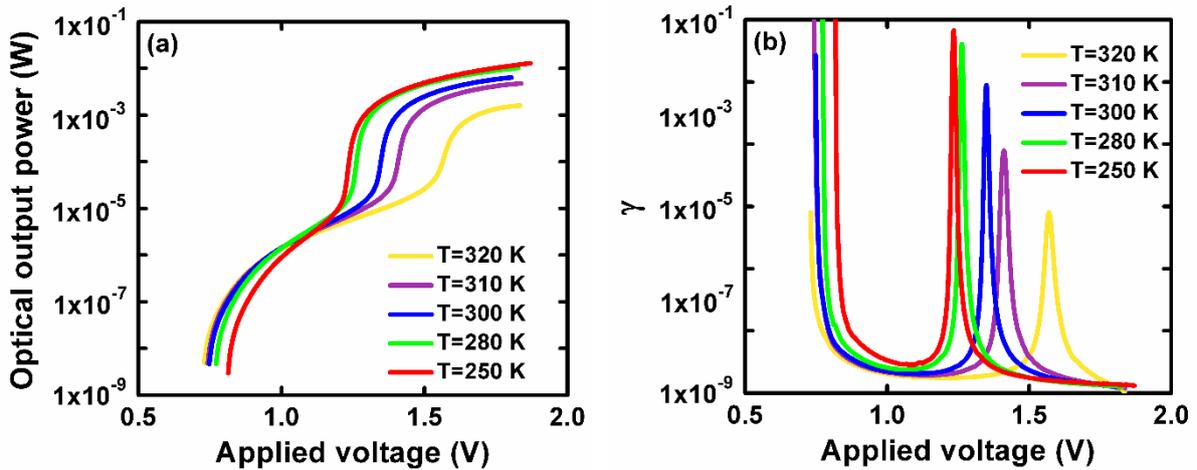

Fig. 10. Temperature dependence of the $P-V$ (a) and the corresponding $\gamma-V$ (b) characteristics.

An increase in temperature widens the voltage range where spontaneous emission dominates however, the voltage corresponding to the onset of spontaneous emission is not very temperature sensitive. Elevated temperatures cause a decrease in the intensity of the stimulated emission, measured at constant voltage as well as in the slope of the $P-V$ curves. The peak value of $\gamma$ decreases with temperature just as the height of the discontinuity in the α-V characteristics (see Fig. 2(b)).

The laser output power can be represented as an exponential function of the applied voltage $P = P_0 V^\gamma$ similarly to the dependence of the current $I = I_S V^\alpha$. In analogy to equations (1) and (2), the output power can be described as:

$$P = P_0 \exp\left(\frac{q V_{tot,p}}{\xi kT}\right), \qquad (13)$$

Where

$$\xi = n\frac{\alpha}{\gamma} \qquad\qquad (14)$$

$\xi$ plays the role of the ideality factor $n$, used to characterize current flow in all diodes. $P_0$ is the pre-exponential coefficient denoting the onset of radiative recombination. $V_{tot,p}$ is the applied voltage at which emission sets in and can generally include voltage drops on any part of the structure that affects the current transport and hence the light intensity dependence on the applied voltage.

The equation is conceptually similar to that proposed in [40]. It differs however in that it takes into account the effect of voltage drops on the nonlinear and or linear resistance similar to the discussion above in relation to the $I - V$ dependencies. The parameters $\xi$ and $R_S$ are found in analogy with equations (8) from the current values of $P, V, \alpha$ and $\gamma$ by using:

$$\xi = \frac{qV}{kT}\frac{(\alpha+\theta-1)}{\gamma(\alpha+\theta-\nu)}; \ R_S = \frac{V}{I}\frac{(1-\nu)}{\alpha(\alpha+\theta-\nu)} \qquad (15)$$

Where $\nu = \frac{V}{\gamma}\frac{d\gamma}{dV}$. The parameters $\xi$ and $n$ are related by equation (14) for given $\alpha$ and $\gamma$ values.

The experimental $P - V$ and corresponding $\gamma - V$ characteristics, at a given temperature, reveal that the start of illumination occurs at a voltage near the end of the exponential part of the $I - V$ characteristics namely, radiative recombination takes place only in the high injection regime. This issue has not been discussed previously in the literature, although it is extremely important to elucidate the features of the spontaneous and stimulated emission regimes in all laser diodes.

The modified expression of the $P - V$ dependence is:

$$P = B_p(V_{tot,p} - V_{cp})^\kappa \quad (16)$$

Where, $B_P, V_{cp}$ and $\kappa$ are analogous to the parameters in equation (4). Using equation (16), these parameters can be extracted from the experimental $P - V$ and $\gamma - V$ characteristics using:

$$\kappa = -\frac{\gamma}{(\nu-1)}; \quad V_{cp} = \frac{\nu V}{(\nu-1)}; \quad B_p = P\left(-\frac{\nu-1}{V}\right)^{\kappa} \quad (17)$$

It is important to establish a relationship between the illumination intensity, the current density and radiatively recombined carriers in the high injection regime. The combination of equations (4), (11) and (16) makes it possible to obtain the $P - I$ and $P - N$, dependencies formulated as:

$$P = B_p \left[\frac{I}{B_i} - \left(V_{cp} - V_{ci}\right)^m\right]^{\frac{\kappa}{m}} \quad (18)$$

$$P = B_p \left[\left(\frac{N}{CmB_i}\right)^{\frac{m}{(m-1)}} - \left(V_{cp} - V_{ci}\right)^m\right]^{\frac{\kappa}{m}} \quad (19)$$

For a constant contact potential (which means no differences in the second terms of equations (18) and (19), allows to reduce (18) and (19) to:

$$P = B_p \left(\frac{I}{B_i}\right)^{\frac{\kappa}{m}} \quad (20)$$

$$P = B_p \left(\frac{N}{CmB_i}\right)^{\frac{\kappa}{(m-1)}} \quad (21)$$

It follows from equations (20) and (21) that linear and quadratic dependencies of the output power on current and carrier densities are only possible for $\kappa = m = 2$. For other values of $\kappa$ or $m$ the $P - I$ characteristics may be linear when $\kappa = m$ but the $P - N$ behavior is always nonlinear.

Extracted parameters from the $P - V$ data are used to fit the experimental characteristics in the entire forward bias range by employing equations (16) and (17). The fitting results are shown in Fig 11. Solid lines represent simulated $P - V$ and $\gamma - V$ curves at 250 K, 280 K and 320 K while broken lines describe the experimental data, taken from Fig. 10. The superb fits confirm the effectiveness of the approximation (16) to describe the $P - V$ characteristics in both regimes of radiative emission.

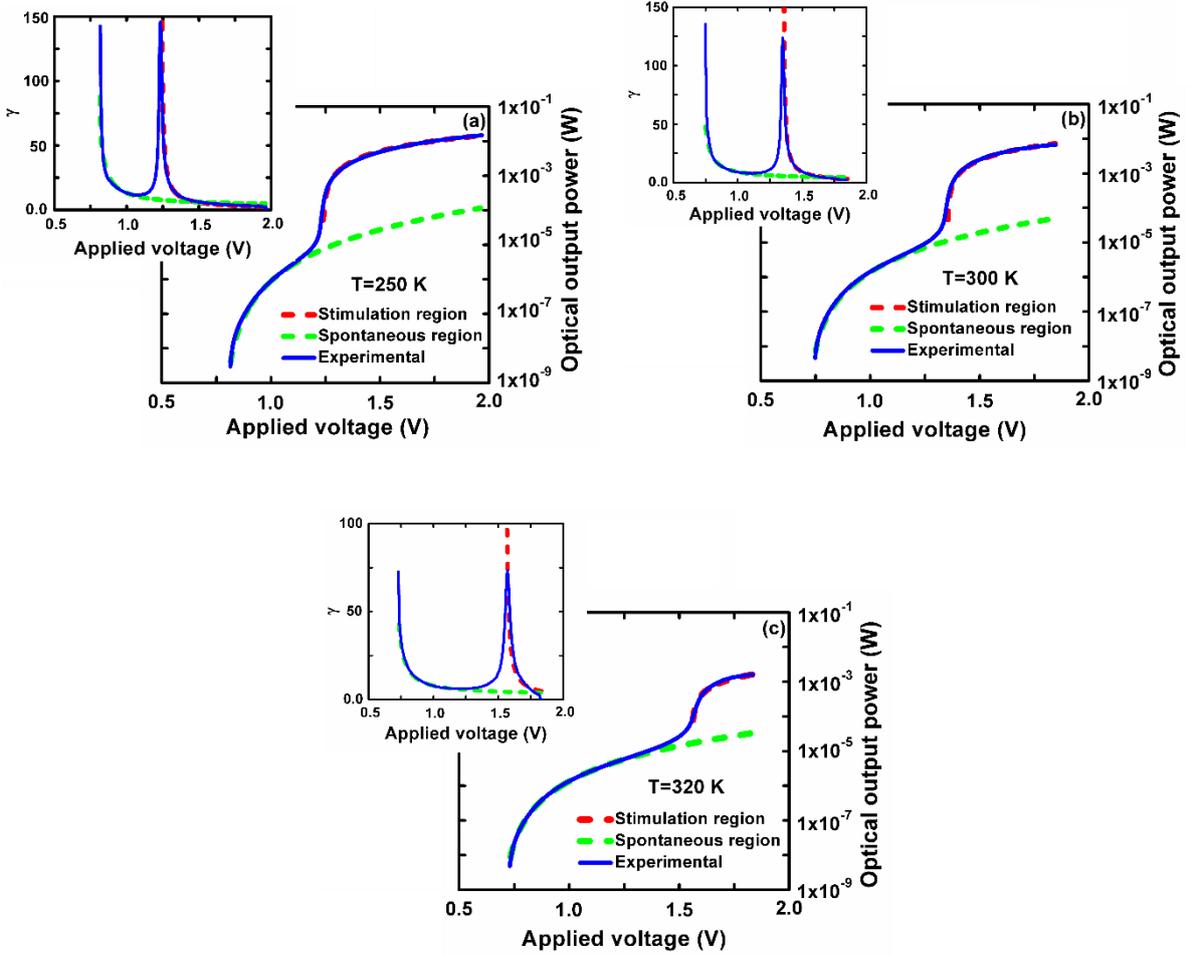

Fig. 11. Experimental (solid lines) and simulated (dash lines) $P - V$ and $\gamma - V$ characteristics. (a) T=250 K, (b) T=280 K and (c) T=320 K.

The temperature dependencies of the characteristic voltages (the second equation in (17)) corresponding to the onset of spontaneous and stimulated emission are shown in Fig. 12 (a). For the spontaneous emission, $V_{cp}$ is linear and when extrapolated to $T = 0\ K$ it yields $V_{cp} = 0.97\ eV$, which is exactly equal to the value, obtained from the Arrhenius plot for the saturation current, of band gap energy of the optical confinement layer discussed above and in [12]. The extracted contact potentials are very close to the values obtained from the $I - V$ characteristic in the high injection mode (see Fig. 7), and hence extrapolation to zero temperature yields $V_{cp}$ which is the energy of the band gap that is extracted from the temperature dependence $V_{ci}$ (see Fig. 7). The stimulated regime is highly nonlinear with a minimum near 250 K and a symmetric increase below and above 250 K.

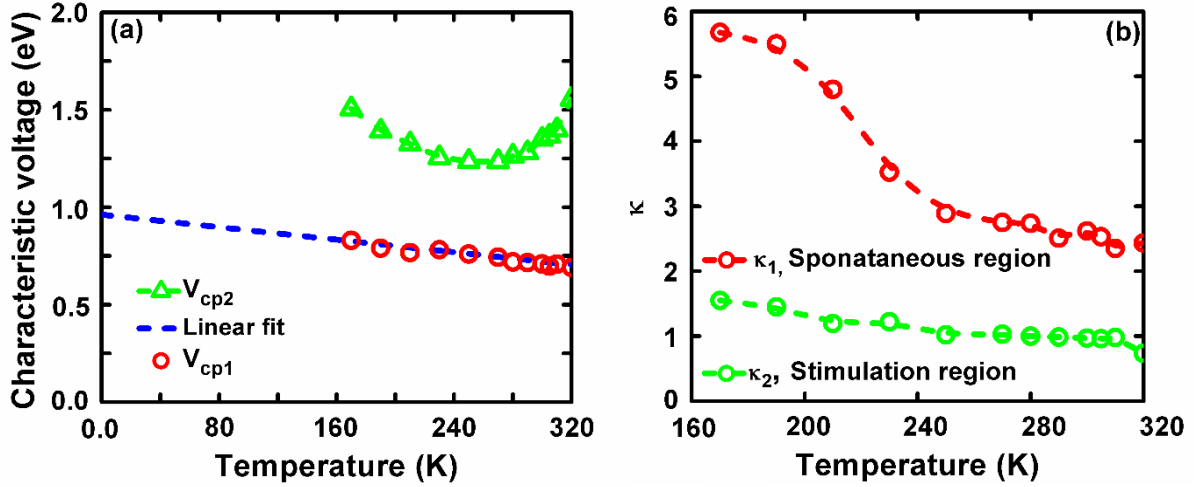

Fig. 12. (a) Temperature dependencies of the forward voltage drop on junctions and set-in voltage. (b) Power exponent $\kappa$ in the spontaneous and stimulation emission regimes.

This result confirms once more the validity of approximation (16) for describing $P-V$ in the high injection mode in $P-I-N$ diodes as predicted in [23] and demonstrated in [12, 26, 27,] for the current flow in $P-I-N$ diodes operating in a similar regime.

For the stimulated emission above the minimum, a voltage increase across the active region mandates a current increase in order to maintain a high level of carrier injection. Below the minimum, a bias increase in needed in order to compensate for the decrease in thermally generated carriers at a reduced temperature. Low temperature causes carrier freeze-out, an increase of the semiconductor resistivity and hence the differential resistance of the OCL region, which causes in turn, a redistribution of the voltage drops across it. An additional voltage is also required to allow carriers to overcome the potential barriers between the device sections, which is due to the depletion width of the diode hetero barriers which include several layers with different forbidden gaps, [40]. Exactly the same curves are obtained when the voltages are extracted from the discontinuity of the α - V curves corresponding to the threshold currents. Finally, as the temperatures decreases, the freeze out of excited-state dopant carriers results in a reduced electrical conductivity in the bulk as well as a further depletion at the epitaxy hetero barrier [41].

A decrease in the power exponents $\kappa$ (equation (16)) is revealed for both radiative recombination regimes as shown in Fig. 12 (b). For spontaneous emission, $\kappa$ decrease by a factor larger than 2 while for stimulated emission, $\kappa$ decreases by a factor of only 1.5. In the temperature range between 250 K to 310 K, both curves saturate at $\kappa$ values of 2.5 and 1. This trend, mainly for the spontaneous emission, is similar to the dependence on temperature of the ideality factor stemming from tunneling in the $I-V$ characteristics. We postulate that the thermally activated process of tunneling injection of carriers plays a role and also contributes to the high injection regime where spontaneous emission sets in. Note, that behavior of the parameter $\kappa$ with temperature is similar to that is observed for $m$ shown in Fig. 7.

### 3. Optical output power current characteristics

In this section, we address the more conventional, optical output power dependence on current. Standard $P-I$ and $\beta-I$ characteristics measured at different temperatures are shown in Fig. 13. The threshold current $I_{th}$ changes from 16.7 mA to 70 mA as the temperature rises from 210 K to 320 K as seen in Fig. 12 (a) and (b).

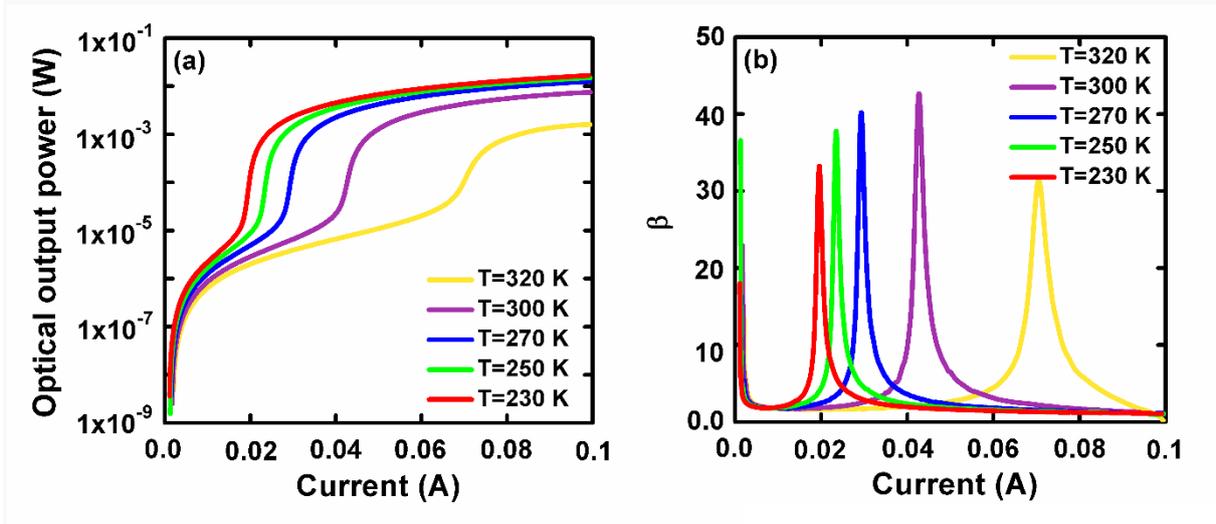

Fig. 13. Experimental (a) P- I and (b) $\beta-I$ characteristics of TI QD laser measured at different temperatures.

Temperature dependent changes can be clearly traced on the $\beta-I$ characteristics shown in Fig. 13 (b), which exhibit, at each temperature, three distinct regions. The first represents spontaneous emission where $\beta$ reduces from about 1.8 to 1.6 with a temperature increase from 210 K to 320. The second regime shows a sharp maximum at the laser threshold current. The third regime describes $\beta$ beyond threshold (where stimulation emission dominates) which reduces continuously with current. Once the thermal roll-over region is reached, $\beta$ reduces to below 1 namely, the characteristics becomes sub-linear.

The $P-I$ and $\beta-I$ curves were simulated using parameters extracted from the experimental data of $I-V$ and $P-V$ by employing equations (4) (16) and (20). Fig. 14 shows the comparisons for 250 K, 300 K and 320 K where broken lines are simulated results and solid lines describe measured data. Extracted parameters from the $P-V$ curves allow fitting the experimental characteristics in the entire forward bias range by employing equations (16) and (17). The superb fittings in both regimes of radiative recombination validate equation (20).

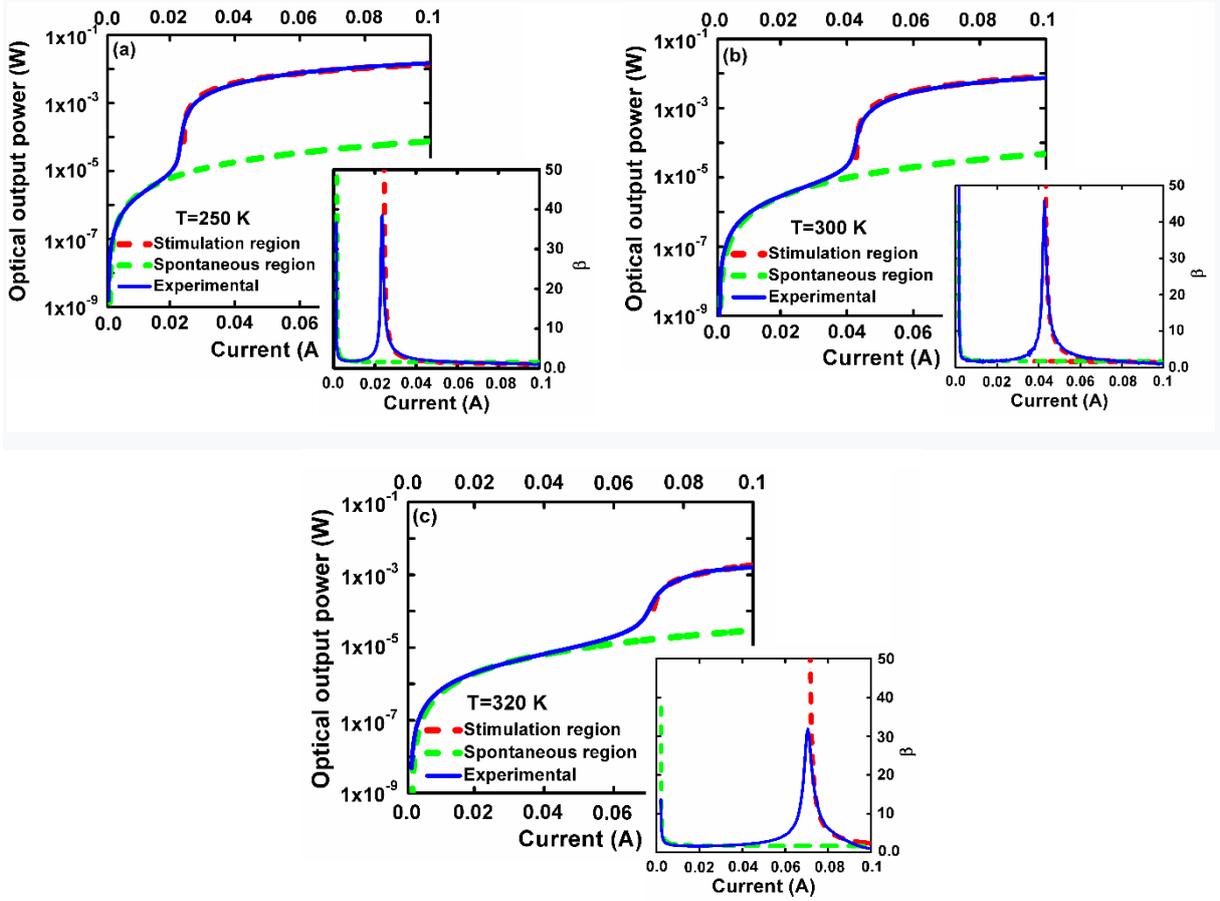

Fig. 14. Experimental (solid lines) and simulated (dash lines) $P-I$ and $\beta-I$ characteristics of TI QD laser. (a) T=250 K, (b) T=300 K and (c) T=320 K.

Using equations (11) and (21) it is possible to established accurate relationships between the optical output power, the drive current and the actual carrier density involved in the radiative recombination. Figure 15 (a) describes the temperature dependencies of the power exponent $m/(m-1)$, characterizing the mutual dependence of the recombination current and the carrier density, in accordance with equation (11) with the parameter $m$ extracted by equation (4) from the measured I – V characteristics in the high injection regime.

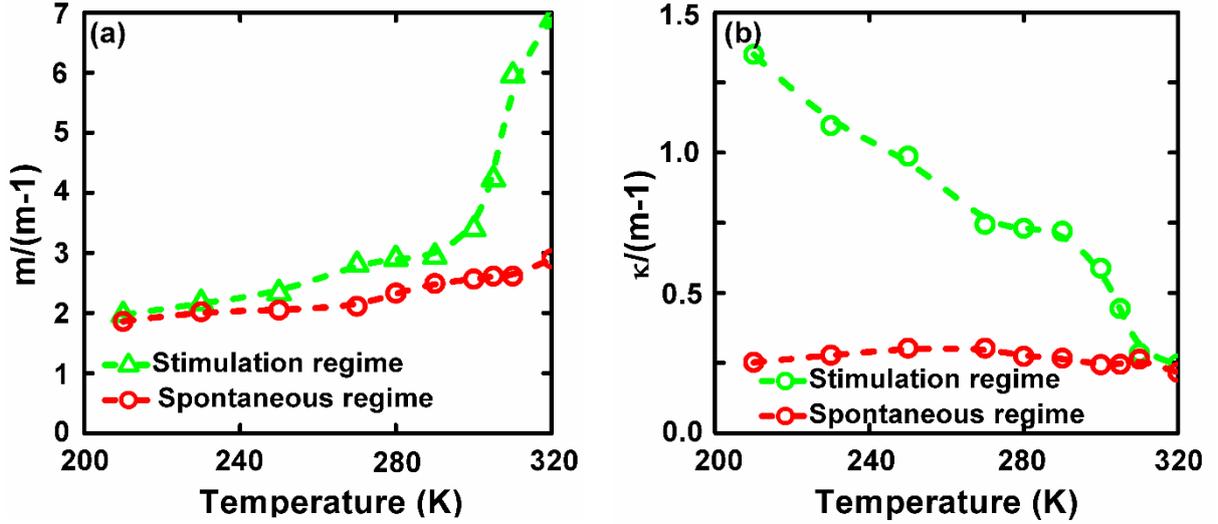

Fig. 15. Temperature dependencee of the power exponents (a) $m/(m-1)$ and (b) $\frac{\kappa}{m-1}$ defining the relationships between the optical output power of a TI QD laser, the current and the carrier density.

Fig. 15 (b) shows a similar functional dependence on of the power exponent $\frac{\kappa}{m-1}$ defining the variation of the optical output power with carrier density (see equation (21)). The change in $m/(m-1)$ is roughly consistent with the standard formalism $I \sim AN + BN^2 + CN^3 + DN^\vartheta$, assuming simple Boltzmann statistics [42-44]. $N = N_i \exp\left(\frac{qV}{nkT}\right)$ is the product of hole and electron densities, while $N_i$ is an intrinsic carrier density. Each term in the $ABCD$ model has a different contribution to the total $I-V$ characteristic resulting from particular recombination processes with distinct ideality factors. The coefficient $A$ corresponds to weak radiative monomolecular recombination in accordance with SHR mechanism with $n = 2$, $B$ relates to bimolecular radiative recombination, $n = 1$ and corresponds to radiative recombination. C describes the Auger recombination process with $n = 2/3$. D and $\vartheta$ was introduced in [43] as parameters defining a sharp nonradiative loss processes due to leakage of carriers through the barrier regions of devices.

Fig 15 (a) shows that for 210 K to 270 K, in the spontaneous regime, the current depends on carrier density in an almost quadratic manner as predicted in the ABC model. In the stimulated regime however, this holds only up to 250 K. Between 250 K and 320 K, for the first radiation regime, the exponential factor increases slowly to roughly 3, but for the stimulated regime, this value increases sharply and reaches the value of 7 at 320 K. This trend is similar to that predicted in [43] but behaves oppositely with temperature. The large power exponent $m/(m - 1)$ during stimulated emission observed above 285 K is related to a carrier leakage process. A second reason is a reduction of the radiation recombination rate due to an increased carrier escape rate at high temperatures. Both processes are less significant when $(m/(m - 1))$ values vary between 2 and 3 since lower currents cause less junction heating.

The curves plotted in Fig. 15 (b) demonstrate the distinct difference between the $P - N$ and the $I - N$ dependencies in both emission modes. In the spontaneous emission regime, optical output power density is strongly sublinear with respect to the carrier density and it also depends weakly on temperature. In contrast, in the stimulated emission regime, the values of the power exponents are small at high temperatures (they are almost equal to those in the spontaneous emission mode) but they increase sharply as the temperature reduces. This means that the injected carriers density grows exponentially in the active region with a temperature reduction so that the part of the fraction of carriers that recombine radiatively increase, as revealed in values of $\kappa$, (see equations (4a), (16) and (11), (21).

The temperature dependence of the threshold current, plotted in a logarithmic scale is shown in Fig. 16. Two linear regions with different slope are clearly seen which allow to estimate the characteristic temperature $T_0$, that is the measure of the of temperature sensitivity of physical parameters of the laser diodes. Using the empirical relation $I_{th} = I_{th0} \exp{(\frac{T}{T_0})}$ [45] yields $T_0$ values of 102 K in the range of 210 K to 280 K and 53.2 K for 280 K to 320 K. The significant reduction in $T_0$ is most likely due to junction heating at elevated ambient temperatures [46]. A different result is obtained at temperature above 280 K, evidently due to enhancement of the junction temperature at room temperature and above [46].

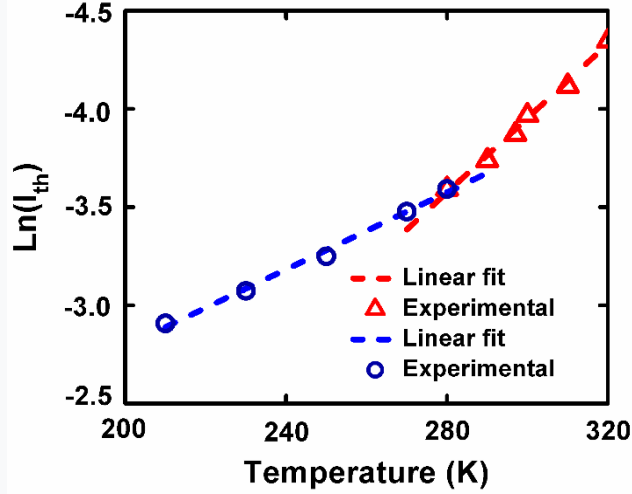

Fig. 16. Temperature dependence of the threshold current of the TI QD laser.

## IV. Conclusion

In conclusion, we investigated the electrical and electro-optical characteristics of a TI QD laser with six *InAs* QDs separated by *InAlGaAs* barriers and *InGaAs* tunnelling layers, in a wide range of current, applied voltage and temperatures. Bias dependent current flow mechanisms are revealed before the onset of radiation. A thermally activated branch of the $I - V$ characteristic was identified at the lowest bias regime in which a diffusion-recombination mechanism dominates. The tunneling process is identified as the dominant carrier transport mechanism at a somewhat increased voltage, corresponding to the second and third exponential parts, due to the strong and weak temperature dependence respectively of the ideality factor and slope of the $I - V$ characteristic, as well as the extracted saturation currents. The extraction of parameters, the behavior of which as a function of temperature is usually the identifying signature of the dominant of carrier flow mechanisms, was found taking into account the influence of high carrier injection into the internal active region of the device, as well as the effect of a series resistance caused by the bulk and the electrodes. This approach is fundamentally different from the common approaches in the SRH model. For these purposes, an analytical formalism was developed, which makes it possible to more accurately simulate the measured dependencies by the extracted parameters. Comparing the TI QD lasers with a QD laser having nominally the same structure, revealed naturally an absence of the tunneling component in the $I - V$ characteristic the QD laser. Finally, an analysis of the optoelectronic characteristics based on a similar analytical formalization of the $I - V$ characteristic showed that they are informative not only from the point of view of establishing the radiation mechanism, but also for identifying some physical parameters (contact potential and band gap of the active region), which are identical to those found from the $I - V$ characteristics. This testifies to the reliability of the predicted mechanisms based on electrical properties. In addition, this formalism establishes a direct relationship between the values of the output power of light and current, as well as the applied voltage and, hence, the injected density of carriers involved in the radiative recombination processes.